\documentclass[pdflatex,prl,twocolumn,showpacs,superscriptaddress,nofootinbib]{revtex4-1}
\usepackage{bm,amsmath,amssymb}
\usepackage{graphicx}
\usepackage{subfigure}
\usepackage{color}
\usepackage{soul}
\usepackage{hyperref}
\usepackage{multirow}
\usepackage{lineno}
\setulcolor{blue}
\setstcolor{red}
\sethlcolor{yellow}

\newcommand \ie {{\sl i.e.}}

\newcommand \etal {{\sl et al.}}

\newcommand \beq {\begin{equation}}
\newcommand \eeq {\end{equation}}
\newcommand \beqa {\begin{eqnarray}}
\newcommand \eeqa {\end{eqnarray}}

\newcommand \hmu {\hat{\mu}}
\begin{document}

\title{Additional Strange Hadrons from QCD Thermodynamics and Strangeness Freeze-out
in Heavy Ion Collisions}

\author{A. Bazavov}
\affiliation{Department of Physics and Astronomy, University of Iowa, Iowa City, Iowa
52240, USA}
\author{H.-T. Ding}
\affiliation{ Key Laboratory of Quark \& Lepton Physics (MOE) and Institute of
Particle Physics, Central China Normal University, Wuhan 430079, China}
\author{P. Hegde}
\affiliation{ Key Laboratory of Quark \& Lepton Physics (MOE) and Institute of
Particle Physics, Central China Normal University, Wuhan 430079, China}
\author{O. Kaczmarek}
\affiliation{Fakult\"at f\"ur Physik, Universit\"at Bielefeld, D-33615 Bielefeld,
Germany}
\author{F. Karsch}
\affiliation{Fakult\"at f\"ur Physik, Universit\"at Bielefeld, D-33615 Bielefeld,
Germany}
\affiliation{Physics Department, Brookhaven National Laboratory, Upton, New York 11973, USA}
\author{E. Laermann}
\affiliation{Fakult\"at f\"ur Physik, Universit\"at Bielefeld, D-33615 Bielefeld,
Germany}
\author{Y. Maezawa}
\affiliation{Fakult\"at f\"ur Physik, Universit\"at Bielefeld, D-33615 Bielefeld,
Germany}
\author{Swagato Mukherjee}
\affiliation{Physics Department, Brookhaven National Laboratory, Upton, New York 11973, USA}
\author{H. Ohno}
\affiliation{Physics Department, Brookhaven National Laboratory, Upton, New York 11973, USA}
\affiliation{Center for Computational Sciences, University of Tsukuba, Tsukuba,
Ibaraki 305-8577, Japan}
\author{P. Petreczky}
\affiliation{Physics Department, Brookhaven National Laboratory, Upton, New York 11973, USA}
\author{C. Schmidt}
\affiliation{Fakult\"at f\"ur Physik, Universit\"at Bielefeld, D-33615 Bielefeld,
Germany}
\author{S. Sharma}
\affiliation{Fakult\"at f\"ur Physik, Universit\"at Bielefeld, D-33615 Bielefeld,
Germany}
\author{W. Soeldner}
\affiliation{Institut f\"ur Theoretische Physik, Universit\"at Regensburg, D-93040
Regensburg, Germany}
\author{M. Wagner}
\affiliation{Physics Department, Indiana University, Bloomington, Indiana 47405, USA}

\begin{abstract}

We compare lattice QCD results for appropriate combinations of net strangeness
fluctuations and their correlations with net baryon number fluctuations with
predictions from two hadron resonance gas (HRG) models having different strange
hadron content. The conventionally used HRG model based on experimentally established
strange hadrons fails to describe the lattice QCD results in the hadronic phase close
to the QCD crossover. Supplementing the conventional HRG with additional,
experimentally uncharted strange hadrons predicted by quark model calculations and
observed in lattice QCD spectrum calculations leads to good descriptions of strange
hadron thermodynamics below the QCD crossover. We show that the thermodynamic
presence of these additional states gets imprinted in the yields of the ground-state
strange hadrons leading to a systematic 5--8 MeV decrease of the chemical freeze-out
temperatures of ground-state strange baryons.

\end{abstract}

\pacs{11.10.Wx, 11.15.Ha, 12.38.Gc, 12.38.Mh}

\maketitle


\emph{Introduction.---} With increasing temperature the strong interaction among
constituents of ordinary nuclear matter, mesons and baryons, results in the copious
production of new hadronic resonances.  The newly produced resonances account for the
interaction among hadrons to an extent that bulk thermodynamic properties become well
described by a gas of uncorrelated hadronic resonances \cite{Hagedorn}. The hadron
resonance gas (HRG) model is extremely successful in describing the hot hadronic
matter created in heavy ion experiments \cite{PBM}.  Abundances of various hadron
species measured in heavy ion experiments at different beam energies are well
described by thermal distributions characterized by a freeze-out temperature and a
set of chemical potentials $\vec{\mu}=(\mu_B,\mu_Q,\mu_S)$ for net baryon number
($B)$, electric charge ($Q$) and strangeness ($S$) \cite{freeze}. Nonetheless,
details of the freeze-out pattern may provide evidence for a more complex sequential
freeze-out pattern \cite{Satz,Heinz}. In particular, in the case of strange hadrons
arguments have been put forward in favor of a greater freeze-out temperature than
that of nonstrange hadrons \cite{Bellwied,Gupta,Bugaev}. 

At the temperature $T_c=(154 \pm 9)$ MeV \cite{hotQCDTc} strong interaction matter
undergoes a chiral crossover to a new phase. In the same crossover region HRG-based
descriptions of the fluctuations and correlations of conserved charges for light
\cite{Redlich}, strange \cite{strange,Bellwied}, as well as charm \cite{charm}
degrees of freedom break down. Below $T_c$, bulk thermodynamic properties as well as
conserved charge distributions are generally well described by a HRG containing all
experimentally observed resonances (PDG-HRG) listed in the particle data tables
\cite{PDG}.  However, there are also some notable differences between lattice QCD
results and the PDG-HRG predictions. At temperatures below $T_c$, the trace anomaly,
\ie, the difference between energy density and 3 times the pressure, is found to be
greater in lattice QCD calculations \cite{fodoreos,fodoreosupdate,hotQCDneweos} than
that of the PDG-HRG. Fluctuations of net strangeness and correlations between net
strangeness and baryon number fluctuations are also larger in QCD compared to those
of PDG-HRG \cite{fodorHRG,hotQCDHRG}. It has been argued that the former provides
evidence for contributions of additional, experimentally still-unobserved hadron
resonances to the thermodynamics of strong interaction matter \cite{Majumder,
NoronhaHostler:2009cf}. Indeed a large number of additional resonances has been
identified in lattice QCD \cite{LGT} and quark model calculations \cite{Isgur,Ebert}.
The presence of such additional flavored hadrons in a thermal medium enhances
fluctuations of the associated quantum numbers and modifies their correlations with
other quantum numbers. In fact, lattice QCD results on net charm fluctuations and
their correlations with baryon number, electric charge, and strangeness fluctuations
have also been compared with the expectations from a HRG containing additional,
experimentally unobserved charmed hadrons predicted by quark model calculations
\cite{charm}. Such comparisons have provided evidence for the thermodynamic
importance of additional charmed hadrons in the vicinity of the QCD crossover
\cite{charm}.

In this Letter, we show that discrepancies between lattice QCD results and PDG-HRG
predictions for strangeness fluctuations and correlations below the QCD crossover can
be quantitatively explained through the inclusion of additional, experimentally
unobserved strange hadrons. The thermodynamic presence of these additional strange
hadrons also gets imprinted on the yields of ground-state strange baryons, resulting
in observable consequences for the chemical freeze-out of strangeness in heavy ion
collision experiments.  

 
\begin{figure}[t]
\includegraphics[scale=0.6]{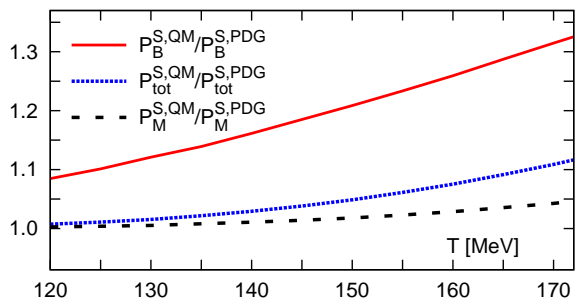}
\caption{Ratios of partial pressures of open strange hadrons
($P^{S,X}_{\mathrm{tot}}$), mesons ($P_M^{S,X}$) and baryons ($P_B^{S,X}$) calculated
in HRG with particle spectra from the particle data table, $X=PDG$, and with
additional resonances predicted by the relativistic quark model, $X=QM$, respectively
(see text).}
\label{fig:QM_PDG}
\end{figure}

\emph{Hadron resonance gas models.---} The partial pressure of all open strange
hadrons can be separated into mesonic and baryonic components,
$P^{S,X}_{\mathrm{tot}} = P_{M}^{S,X}+P_{B}^{S,X}$, 
\beqa
P_{M/B}^{S,X}(T,\vec{\mu}) = {T^4\over {2\pi^2}}
&& \sum_{i\in X} g_i \left(\frac{m_i}{T}\right)^2 K_2({{m_i/T}}) 
\nonumber \\
&& \times \cosh \left( B_i \hat{\mu}_B + Q_i \hat{\mu}_Q +S_i\hat{\mu}_S \right)
\;. 
\label{Cpressure}
\eeqa
Here, $M$ ($B$) labels the partial pressure of open strange mesons (baryons), $g_i$
is the degeneracy factor for hadrons of mass $m_i$, and $\hmu_q\equiv \mu_q/T$, with
$q=B,\ Q,\ S$. The sum is taken over all open strange mesons or baryons listed in the
particle data tables (X=PDG) or a larger set including additional open strange mesons
\cite{Ebert} and baryons \cite{Isgur} predicted by quark models (X=QM).  Throughout
this work we refer to the HRG model containing these additional, quark model
predicted, experimentally undiscovered hadrons as the QM-HRG. In
Eq.~(\ref{Cpressure}) the classical, Boltzmann approximation has been used which is
known to be appropriate for all strange hadrons at temperatures $T\lesssim T_c$
\cite{strange}.

The masses and, more importantly, the number of additional states are quite similar
in the quark model calculations and the strange hadron spectrum of lattice QCD
\cite{LGT}. HRG models based on either one, thus, give very similar results. As the
lattice computations of the strange hadron spectrum have so far been performed with
unphysically heavy up and down quark masses, for definiteness we have chosen to
compare our finite temperature lattice results with the quark model predictions that
generally reproduce the masses of the experimentally known states rather well.

Figure \ref{fig:QM_PDG} compares partial pressures of open strange mesons and
baryons calculated within PDG-HRG and QM-HRG models. The additional strange baryons
present in the QM-HRG lead to a large enhancement of the partial baryonic pressure
relative to that obtained from the PDG-HRG model. In the mesonic sector, changes are
below 5\% even at $T= 170$~MeV, \ie,  above the applicability range of any HRG
\cite{strange}.  This simply reflects that a large part of the open strange mesons is
accounted for in the PDG-HRG model, and the additional strange mesons contributing to
the QM-HRG model are too heavy to alter the pressure significantly. 


\emph{Strangeness fluctuations and correlations.---} We calculate cumulants of
strangeness fluctuations and their correlations with baryon number and electric
charge in (2+1)-flavor QCD using the highly improved staggered quark (HISQ) action
\cite{Follana}. In these calculations the strange quark mass ($m_s$) is tuned to its
physical value and the masses of degenerate up and down quarks have been fixed to
$m_l =m_s/20$.  In the continuum limit, the latter corresponds to a pion mass of
about 160~MeV. In the relevant temperature range, $145~{\rm MeV} \le T\le 170~{\rm
MeV}$, we have analyzed (10-16)$\times10^3$ configurations, separated by 10 time
units in rational hybrid Monte Carlo updates, on lattices of size $6\times24^3$ and
$8\times32^3$. Some additional calculations on $8\times32^3$ lattices have been
performed with physical light quarks, $m_l=m_s/27$, to confirm that the quark mass
dependence of observables of interest is indeed small. 

To analyze the composition of the thermal medium in terms of the quantum numbers of
the effective degrees of freedom, we consider generalized susceptibilities of the
conserved charges,
\beq
\chi_{klm}^{BQS} = \left.
\frac{\partial^{(k+l+m)} [P(T,\hmu_B,\hmu_Q,\hmu_S)/T^4]}
{\partial \hmu_B^k \partial \hmu_Q^l \partial \hmu_S^m }
\right|_{\vec{\mu}=0}
\;,
\label{eq:susc}
\eeq
where $P$ denotes the total pressure of the hot medium. For brevity, we drop the
superscript when the corresponding subscript is zero.

\begin{figure}[!t]
\includegraphics[scale=0.6]{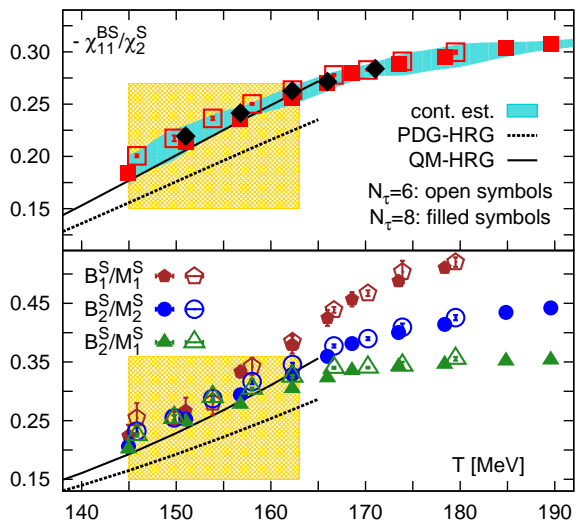}
\caption{Top: $BS$ correlations normalized to the second cumulant of net strangeness
fluctuations. Results are from (2+1)-flavor lattice QCD calculations performed with a
strange to light quark mass ratio $m_s/m_l=20$ (squares) and $m_s/m_l=27$
(diamonds). The band depicts the improved estimate for the continuum result
facilitated by the high statistics $N_\tau=6$ and $8$ data. Bottom: Ratios of
baryonic ($B^S_i$) to mesonic ($M^S_i$) pressure observables defined in
Eq. (\ref{BS_MS}). The dotted and solid lines show results from PDG-HRG and QM-HRG model
calculations, respectively. The shaded region denotes the continuum extrapolated
chiral crossover temperature at physical values of quark masses $T_c=(154\pm
9)$~MeV
\cite{hotQCDTc}.}
\label{fig:BS_S2}
\end{figure}

\begin{figure}[!t]
\includegraphics[scale=0.6]{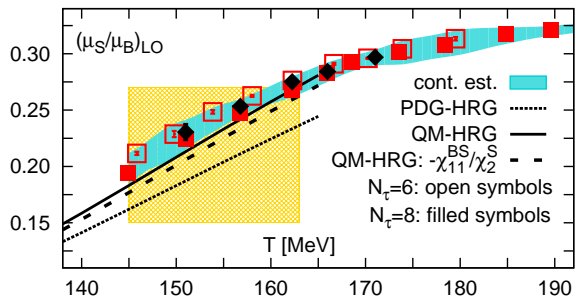}
\caption{The leading-order result for the ratio of strangeness and baryon chemical
potentials versus temperature. Data and HRG model results are for a strangeness
neutral thermal system having a ratio of net electric charge to net baryon number
density $N_Q/N_B =0.4$. The dashed line shows the QM-HRG result for vanishing
electric charge chemical potential.  All other curves and labels are as in
Fig.~\ref{fig:BS_S2}.}
\label{fig:muS_muB}
\end{figure}

The correlation of net strangeness with net baryon number fluctuations normalized to
the second cumulant of net strangeness fluctuations, $\chi_{11}^{BS}/\chi_2^S$ is a
sensitive probe of the strangeness carrying degrees of freedom \cite{Koch}.
Consistent continuum extrapolations for this ratio have been obtained with two
different staggered fermion discretization schemes \cite{fodorHRG,hotQCDHRG}.  In
Fig.~\ref{fig:BS_S2}(top) we show our present, refined results obtained for lattices
with temporal extents $N_\tau= 6$ and 8, which are in agreement with the earlier
results, together with an improved estimate for the continuum result based on the
enlarged statistics for these lattices. In the crossover region and also at lower
temperatures in the hadronic regime the lattice QCD results for
$-\chi_{11}^{BS}/\chi_2^S$ are significantly greater than those of the PDG-HRG model
predictions. 

In the validity range of HRG models, the $BS$ correlation $\chi_{11}^{BS}$ is a
weighted sum of partial pressures of strange baryons, while the quadratic strangeness
fluctuations $\chi_2^S$ are dominated by the contribution from strange mesons. The
larger value of $-\chi_{11}^{BS}/\chi_2^S$ found in lattice QCD calculations compared
to that of PDG-HRG model calculations, thus, reflects the stronger increase of
$P_B^{S,QM}/P_B^{S,PDG}$ compared to $P_M^{S,QM}/P_M^{S,PDG}$ (see
Fig.~\ref{fig:QM_PDG}). As a consequence, the QM-HRG model provides a better
description of QCD thermodynamics in the hadronic phase.  This can be more directly
verified by considering the ratio of two observables, which in a HRG model give
$P_M^{S}$ and $P_B^{S}$, respectively. There is a large set of
``pressure-observables'' that can be constructed for this purpose by using second-
and fourth-order cumulants of strangeness fluctuations and correlations with net
baryon number \cite{strange}.  They will all give identical results in a gas of
uncorrelated hadrons, but differ otherwise. In particular, they can yield widely
different results in a free quark gas.  We use two linearly independent
pressure-observables for the open strange meson $(M^S_1, M^S_2)$ and baryon $(B^S_1,
B^S_2)$ partial pressures, respectively,
\beqa
M^S_1 &=& \chi_2^S -\chi_{22}^{BS} 
\;, \nonumber \\
M^S_2 &=& \frac{1}{12}\left( \chi_4^S +11 \chi_2^S \right)
+ \frac{1}{2}\left( \chi_{11}^{BS} + \chi_{13}^{BS}\right)
\;, \nonumber \\
B^S_1 &=& -\frac{1}{6}\left( 11 \chi_{11}^{BS} + 
6 \chi_{22}^{BS} +\chi_{13}^{BS} \right)
\;, \nonumber \\
B^S_2 &=& \frac{1}{12}\left(\chi_4^S -\chi_2^S\right) - 
\frac{1}{3}\left(4 \chi_{11}^{BS}- \chi_{13}^{BS}\right)
\;.
\label{BS_MS}
\eeqa

Three independent ratios $B^S_i/M^S_j$ are shown in Fig.~\ref{fig:BS_S2}~(bottom).
They start to coincide in the crossover region giving identical results only below $T
\lesssim 155$ MeV. This reconfirms that a description of QCD thermodynamics in terms
of an uncorrelated gas of hadrons is valid till the chiral crossover temperature
$T_c$. Below $T_c$, the information one extracts from $B^S_i/M^S_j$ agrees with that
of $-\chi_{11}^{BS}/\chi_2^S$. In the hadronic regime the ratios calculated in
lattice QCD are significantly greaterthan those calculated in the PDG-HRG model.
QM-HRG model calculations are in good agreement with lattice QCD. These results
provide evidence for the existence of additional strange baryons and their
thermodynamic importance below the QCD crossover.


\emph{Implications for strangeness freeze-out.---} Since the initial nuclei in a
heavy ion collision are net strangeness free, the HRG at the chemical freeze-out must
also be strangeness neutral. Obviously, for such a strangeness neutral medium, all
three thermal parameters $T$, $\mu_B$, and $\mu_S$ are not independent; the
strangeness chemical potential can be expressed as a function of $T$ and $\mu_B$.
While $\mu_S(T,\mu_B)$ is unique in QCD, for a HRG it clearly depends on the relative
abundances of the open strange baryons and mesons. For fixed $T$ and $\mu_B$, a
strangeness neutral HRG having a larger relative abundance of strange baryons over
open strange mesons naturally leads to a larger value of $\mu_S$.

Calculations of $\mu_S(T,\mu_B)$ in a strangeness neutral HRG are straightforward.
For QCD this can be obtained from lattice QCD computations of $\mu_S/\mu_B$ using
next-to-leading-order Taylor expansion of the net strangeness density
\cite{freeze,WBfreeze}. The ratio $\mu_S/\mu_B = s_1(T)+s_3(T) (\mu_B/T)^2 +{\cal
O}(\mu_B^4)$ is closely related to the ratio $\chi_{11}^{BS}/\chi_2^S$ shown in
Fig.~\ref{fig:BS_S2}. At leading order, it only receives a small correction from
nonzero electric charge chemical potential $\mu_Q/\mu_B$,
\beq
\left(\frac{\mu_S}{\mu_B}\right)_{\mathrm{LO}} 
\equiv s_1(T)
= -\frac{\chi_{11}^{BS}}{\chi_2^S}
-\frac{\chi_{11}^{QS}}{\chi_2^S} \frac{\mu_Q}{\mu_B}
\;.
\label{muSmuB}
\eeq
The next-to-leading-order correction $s_3(T) (\mu_B/T)^2$ \cite{freeze} is small for
$\mu_B \lesssim 200$~MeV. We show the leading order result in Fig.~\ref{fig:muS_muB}.
At a given temperature, the strangeness neutrality constraint gives rise to a larger
value, consistent with lattice QCD results, of $\mu_S/\mu_B$ for the QM-HRG compared
to the PDG-HRG. In other words, for a given $\mu_B/T$, the required value of
$\mu_S/\mu_B$ necessary to guarantee strangeness neutrality is achieved at a lower
temperature in the QCD and QM-HRG model than in the PDG-HRG model.

The relative yields of strange anti-baryons ($\bar{H}_S$) to baryons ($H_S$) at
freeze-out are controlled by the freeze-out parameters $(T^f,\mu_B^f,\mu_S^f)$,
\beq
R_H \equiv \frac{\bar{H}_S}{H_S} = 
{\rm e}^{-2(\mu_B^f/T^f)\left(1 - (\mu_S^f/\mu_B^f) |S|\right)}
\; .
\label{HbarH}
\eeq
(For simplicity of the argument we will ignore here the influence of a non-vanishing
electric charge chemical potential. As shown in Fig.~\ref{fig:muS_muB} a non-zero
$\mu_Q/T$ has only a small influence on $\mu_S/\mu_B$.) While this relation does not
explicitly depend on the content and spectra of hadrons in a HRG, the presence of
additional strange hadrons implicitly enters through the strangeness neutrality
constraint. As discussed before, for different HRG models at fixed $T$ and $\mu_B$
strangeness neutrality leads to different values of $\mu_S/\mu_B$. Once $\mu_B^f/T^f$
and $\mu_S^f/\mu_B^f$ are fixed through experimental yields of strange hadrons, it is
obvious from Fig.~\ref{fig:muS_muB} that a given value of $\mu_S^f/\mu_B^f$ is
realized at a larger temperature in the PDG-HRG model than in the QM-HRG model.
 
\begin{figure}[t]
\includegraphics[scale=0.6]{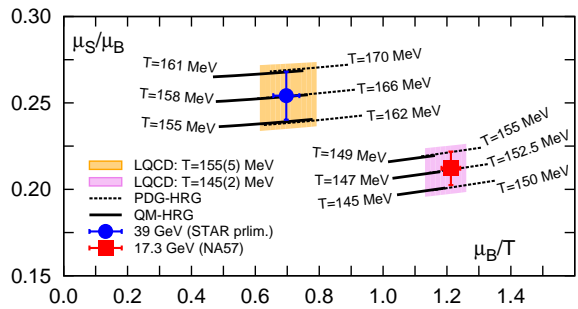}
\caption{Values of ($\mu_S^f/\mu_B^f$,$\mu_B^f/T^f$) extracted from fits to multiple
strange hadrons yields (see text) are compared to $\mu_S/\mu_B$ predictions, obtained
by imposing strangeness neutrality, from lattice QCD calculations (shaded bands) as
well as from QM-HRG (solid lines) and PDG-HRG (dotted lines) models. The predictions
are shown for $\mu_B/T=\mu_B^f/T^f$. For each case, the temperature ranges are chosen
such that the predicted values reproduce $\mu_S^f/\mu_B^f$.}
\label{fig:Tf}
\end{figure}

Ratios of the freeze-out parameters $\mu_B^f/T^f$ and $\mu_S^f/\mu_B^f$ can be
obtained by fitting the experimentally measured values of the strange baryon ratios
$R_\Lambda=\bar{\Lambda}/\Lambda$, $R_\Xi=\Xi^+/\Xi^-$ and $R_\Omega
=\Omega^+/\Omega^-$ to Eq.~\ref{HbarH}. Fits of these strange anti-baryon to baryon
yields to Eq. (\ref{HbarH}) result in $(\mu_S^f/\mu_B^f,\mu_B^f/T^f)=$
(0.213(10),1.213(30)) for the NA57 results at $\sqrt{s}=17.3$ GeV \cite{NA57} and
(0.254(7),0.697(20)) for the  STAR preliminary results at $\sqrt{s}=39$ GeV
\cite{CPOD_STAR}. In Fig.~\ref{fig:Tf} we show comparisons of these
$(\mu_S^f/\mu_B^f,\mu_B^f/T^f)$ values with the lattice QCD, QM-HRG, and PDG-HRG
predictions for $\mu_S/\mu_B$ at $\mu_B/T=\mu_B^f/T^f$.  By varying the temperature
ranges, one can match the values of $\mu_S/\mu_B$ to $\mu_S^f/\mu_B^f$ and, thus,
determine the freeze-out temperatures $T^f$. As expected from Fig.~\ref{fig:muS_muB},
the QM-HRG predictions are in good agreement with lattice QCD results and lead to
almost identical values for $T^f$. The PDG-HRG-based analysis, however, results in
freeze-out temperatures for strange baryons that are larger by about 8 (5) MeV for
the smaller (larger) value of $T^f$.

\emph{Conclusions.---} By comparing lattice QCD results for various observables of
strangeness fluctuations and correlations with predictions from PDG-HRG and QM-HRG
models, we have provided evidence that additional, experimentally unobserved strange
hadrons become thermodynamically relevant in the vicinity of the QCD crossover. We
have also shown that the thermodynamic relevance of these additional strange hadrons
modifies the yields of the ground-state strange hadrons in heavy ion collisions.
This leads to significant reductions in the chemical freeze-out temperature of
strange hadrons. Compared to the PDG-HRG, the QM-HRG model provides a more complete
description of the lattice QCD results on thermodynamics of strange hadrons at
moderate values of the baryon chemical potential. This suggests that the QM-HRG model
is probably the preferable choice for the determination of the freeze-out parameters
also at greater values of the baryon chemical potential beyond the validity of the
present lattice QCD calculations. 

Finally, we note that the observation regarding the thermodynamic relevance of
additional strange hadrons hints that an improved HRG model including further,
unobserved light quark hadrons may resolve the current discrepancy between lattice
QCD results for the trace anomaly and the results obtained within the PDG-HRG model.


\emph{Acknowledgments.---} This work was supported in part through Contract No.
DE-AC02-98CH10886 with the U.S. Department of Energy, through the Scientific
Discovery through Advanced Computing (SciDAC) program funded by the U.S. Department
of Energy, Office of Science, Advanced Scientific Computing Research and Nuclear
Physics, the BMBF under Grants No. 05P12PBCTA and No. 56268409, the DFG under Grant
No. GRK 881, EU under Grant No. 283286 and the GSI BILAER Grant. Numerical
calculations were performed using GPU clusters at JLab, Bielefeld University,
Paderborn University, and Indiana University. We acknowledge the support of Nvidia
through the Cuda Research Center at Bielefeld University.


\end{document}